\documentclass[twocolumn,preprintnumbers,nofootinbib,superscriptaddress]{revtex4}
\usepackage{amsmath} \usepackage{graphicx} \usepackage{amsfonts}
\usepackage{array} \usepackage{amsthm} \usepackage{bm} 

\usepackage{latexsym}

\usepackage[breaklinks]{hyperref}
\usepackage{color}

\newcommand{\nn}{\nonumber}
\newcommand{\be}{\begin{equation}}
\newcommand{\ee}{\end{equation}}
\newcommand{\ba}{\begin{eqnarray}}
\newcommand{\ea}{\end{eqnarray}}
\newcommand{\bal}{\begin{align}}
\newcommand{\eal}{\end{align}}

\newcommand{\e}{{\rm e}}
\newcommand{\dd}{{\rm d}}

\newcommand{\bb}{\bibitem}

\newcommand{\om}{\omega}
\newcommand{\al}{\alpha}

\newcommand{\La}{\Lambda}

\newcommand{\ro}{\rho}
\newcommand{\ep}{\epsilon}

\newcommand{\Ta}{\Theta}

\newcommand{\bw}{\begin{widetext}}
\newcommand{\ew}{\end{widetext}}

\def\ch{u_{\text{ch}}}
\def\eh{u_{\text{eh}}}
\def\ah{u_{\text{ah}}}
\def\rch{r_{\text{ch}}}
\def\reh{r_{\text{eh}}}
\def\rah{r_{\text{ah}}}
\def\teh{T_{\text{eh}}}
\def\Q{quintessence}
\def\q{quintessence }
\def\abh{black hole }
\def\bh{black holes }
\def\aBH{black hole}
\def\BH{black holes}
\def\RN{Reissner-Nordstr\"om }
\def\dS{de Sitter }
\def\Sd{Schwarzschild }

\begin{document}
\title{Charged de Sitter-like black holes: quintessence-dependent enthalpy and new extreme solutions}

\author{Mustapha Azreg-A\"{\i}nou}
\affiliation{Ba\c{s}kent University, Faculty of Engineering, Ba\u{g}l\i ca Campus, 06810 Ankara, Turkey}


\begin{abstract}
We consider Reissner-Nordstr\"{o}m black holes surrounded by quintessence where both a non-extremal event horizon and a cosmological horizon exist besides an inner horizon ($-1\leq \omega <-1/3$). We determine new extreme black hole solutions that generalize the Nariai horizon to asymptotically de Sitter-like solutions for any order relation between the squares of the charge $q^2$ and the mass parameter $M^2$ provided $q^2$ remains smaller than some limit, which is larger than $M^2$. In the limit case $q^2=9\omega^2 M^2/(9\omega^2-1)$, we derive the general expression of the extreme cosmo-black-hole, where the three horizons merge, and discuss some of its properties. We also show that the endpoint of the evaporation process is independent of any order relation between $q^2$ and $M^2$. The Teitelboim's energy and Padmanabhan's energy are related by a nonlinear expression and are shown to correspond to different ensembles. We also determine the enthalpy $H$ of the event horizon, as well as the effective thermodynamic volume which is the conjugate variable of the negative quintessential pressure, and show that in general the mass parameter and the Teitelboim's energy are different from the enthalpy and internal energy; only in the cosmological case, that is, for Reissner-Nordstr\"{o}m-de Sitter black hole we have $H=M$. Generalized Smarr formulas are also derived. It is concluded that the internal energy has a universal expression for all static charged black holes, with possibly a variable mass parameter, but it is not a suitable thermodynamic potential for static-black-hole thermodynamics if $M$ is constant. It is also shown that the Reverse Isoperimetric Inequality holds. We generalize the results to the case of the Reissner-Nordstr\"{o}m-de Sitter black hole surrounded by quintessence with two physical constants yielding two thermodynamic volumes.

\end{abstract}


\maketitle

\section{Introduction\label{sec1}}

The inclusion of the $P$-$V$ term in the first law of thermodynamics or in its familiar equivalent laws~\cite{ex0}-\cite{Mann2} has led to the notion of the effective thermodynamic volume, which is in general different from the geometric volume excluded by, say, the event horizon. The thermodynamic volume is the conjugate variable, with respect to some appropriate thermodynamic potential, of the pressure exerted on the horizon attributable to the presence of a constant cosmological density, or a \Q, or both.

From this point of view, much more progress has been made for anti-de Sitter black holes~\cite{Wu}-\cite{AdS} thanks to the AdS/CFT correspondence, the applicability of which has ever been extended~\cite{Maldacena2}-\cite{Maldacena7}. The dS/CFT emerged as a possible dual relation relating quantum gravity on a de Sitter space to a Euclidean conformal field theory on a boundary of the same space~\cite{CFT1}-\cite{CFT4}. Both these correspondences have motivated the classical and quantum investigations of the de Sitter-like and anti-de Sitter spacetimes.

The inclusion of the $P$-$V$ yields, on the one hand, a generalized Smarr formula preserving a scaling law between thermodynamic variables and, on the other hand, an identification of the mass parameter with the enthalpy of the event horizon. These properties apply to both static and rotating \BH. In the static (non-rotating) case, however, a potential problem exists as noticed by Dolan~\cite{Dolan}: the thermodynamic volume $V$ is a function of the entropy $S$, and conversely, so one of the two variables, $S$ or $V$, is redundant. This implies that the internal energy is not a suitable thermodynamic potential for the thermodynamic description of the static \dS and anti-\dS \BH. When rotation is included, the volume no longer depends on the entropy only, and so it is an independent thermodynamic variable.

In this work we consider Reissner-Nordstr\"om and Reissner-Nordstr\"om-de Sitter black holes surrounded by \q where both a non-extremal event horizon and a cosmological horizon exist besides an inner horizon. These are the asymptotically de Sitter solutions that correspond to $-1\leq \om <-1/3$. The case of asymptotically flat solutions corresponding to $-1/3\leq \om<0$, where only a non-extremal event horizon and an inner horizon exist, was treated in Ref.~\cite{AAR}, so we will not consider it here. As we shall see, some conclusions drawn, and results derived, in this work apply to the case of asymptotically flat solutions too.

In Sec.~\ref{sec2} we briefly review the Reissner-Nordstr\"om black holes surrounded by \q derived in Ref.~\cite{Kis}; some other of their properties are discussed in Ref.~\cite{AAR}.

As is well known, extreme \BH, while instable, are important ingredients in the theory of quantum gravity where one can find a microscopic explanation of the Bekenstein-Hawking entropy~\cite{liv}. Another type of extreme \BH, also instable, known as Nariai-type solutions~\cite{Nariai1}-\cite{Nariai3b} are also used in quantum gravity, where some singularities may be replaced by a Nariai type universes~\cite{Nariai4}, besides their use for generating new solutions~\cite{Nariai5,Nariai6}. Some special Nariai black holes with quintessence have been discussed in~\cite{cold}. In Sec.~\ref{sec3} we will determine explicitly new extreme \abh solutions that generalize the Nariai horizon~\cite{Nariai1,Nariai2} to all asymptotically de Sitter-like solutions ($-1\leq\om <-1/3$) for any order relation between the squares of the electric charge $q^2$ and the mass parameter $M^2$ provided $q^2$ remains smaller than some limit, which depends on $\om$ and remains proportional to, but larger than, $M^2$. In the limit case $q^2=9\om^2 M^2/(9\om^2-1)$, the three horizons merge and we derive the general expression of the extreme cosmo-black-hole and discuss some of its properties.

In the first part of Sec.~\ref{sec4}, we consider the thermodynamics of the event horizon and investigate the evaporation process and its endpoint by providing the final values of the mass parameter and the radius of the event horizon.

The second part of Sec.~\ref{sec4} is devoted to a discussion of the conserved charges, mainly, the energy. Because of the nonexistence of global timelike Killing vector for the de Sitter-like spacetimes, there is no notion of spatially asymptotically conserved charges which is similar to that for asymptotically flat or anti-de Sitter spacetimes. Other notions of conserved charges, however, exist but generally lead to different values of the charges. Using different approaches, some authors~\cite{AD}-\cite{Pad3} were led to simple prescriptions when applied to asymptotically de Sitter-like \BH, among which we will discuss the Teitelboim's energy~\cite{Tt1,Tt2} and the Padmanabhan's energy~\cite{Pad1,Pad2,Pad3}. We will relate these two notions of energy and show that they correspond to different ensembles. This will be clarified noticing, beforehand, that the notion of ensembles for the de Sitter-like spacetimes is larger than that of classical thermodynamics.

In Sec.~\ref{sec5} we will determine the enthalpy $H$ of the event horizon, as well as the effective thermodynamic volume, and show that in general the mass parameter and the Teitelboim's energy are different from the enthalpy; only in the cosmological case $\om=-1$, that is, for Reissner-Nordstr\"om-de Sitter \abh we have $H=M$. Generalized Smarr formulas are also derived. It is concluded that the internal energy is not a suitable thermodynamic potential for the thermodynamic description of the static de Sitter-like \BH.

In Sec.~\ref{sec5b} we generalize the results concerning the thermodynamics to the case of the Reissner-Nordstr\"om-de Sitter black hole surrounded by quintessence with two physical constants. We conclude in Sec.~\ref{sec6}.

\section{\RN \bh surrounded by \q \label{sec2}}

In 4-dimensional spacetime, a spherical symmetric \RN \abh plunged into the field of a spherical symmetric \q has the metric~\cite{Kis,AAR}
\begin{equation}\label{2.1}
    \dd s^2=f(r)\dd t^2-f^{-1}(r)\dd r^2-r^2\dd \Omega^2
\end{equation}
with
\begin{equation}\label{2.2}
f(r)=1-\frac{2M}{r}+\frac{q^2}{r^2}-\frac{2c}{r^{3\om+1}}, -1\leq \om<0 \text{ and } c>0.
\end{equation}
With this notation and the convention $G=\hbar=1$, the density of energy and isotropic pressure of \q are
\begin{equation}\label{2.3}
    \ro_{\text{q}}=-\frac{3\om c}{4\pi r^{3\om+3}}>0,\quad p_{\text{q}}=\om \ro_{\text{q}}<0.
\end{equation}
Here and in Ref.~\cite{AAR} $c$ is half the opposite of its original value~\cite{Kis}. The convention used in Ref.~\cite{Kis} is such that $4\pi G=1$ where the expressions of ($\ro_{\text{q}},p_{\text{q}}$) have different constant coefficients. This same convention was used in Refs.~\cite{cold}-\cite{SF} and partly in Ref.~\cite{AAR}.

The \bh described by~\eqref{2.1} and~\eqref{2.2} are classified according to their asymptotic behavior~\cite{AAR}
\begin{align}
\label{2.3a}&\hspace{-0.5mm}\tfrac{-1}{3}\leq \om<0: \text{ asymptotically flat solutions}\\
\label{2.3b}&\hspace{-0.5mm}-1\leq \om <\tfrac{-1}{3}: \text{ asymptotically de Sitter solutions},
\end{align}
with different physical properties depending on the sign of $3\om+1$.  We worked out the case of asymptotically flat solutions in~\cite{AAR}. In this work we shall consider the asymptotically de Sitter solutions. This corresponds to $-2\leq 3\om+1 < 0$ ($-1\leq\om <-1/3$). This will extend the work done in~\cite{AAR}, which necessitated a special treatment different from the one we are aiming to perform here, to asymptotically de Sitter solutions.

Not all solutions are tractable analytically. In Sect.~\ref{sec3}, we will keep doing general treatments and we will  deal particularly with the cosmological constant case $\om =-1$ ($3\om+1=-2$), which is the \RN \abh in the de Sitter space with $\La =6c$, and the case $\om =-2/3$ ($3\om+1=-1$).

\begin{figure*}
\centering
  \includegraphics[width=0.32\textwidth]{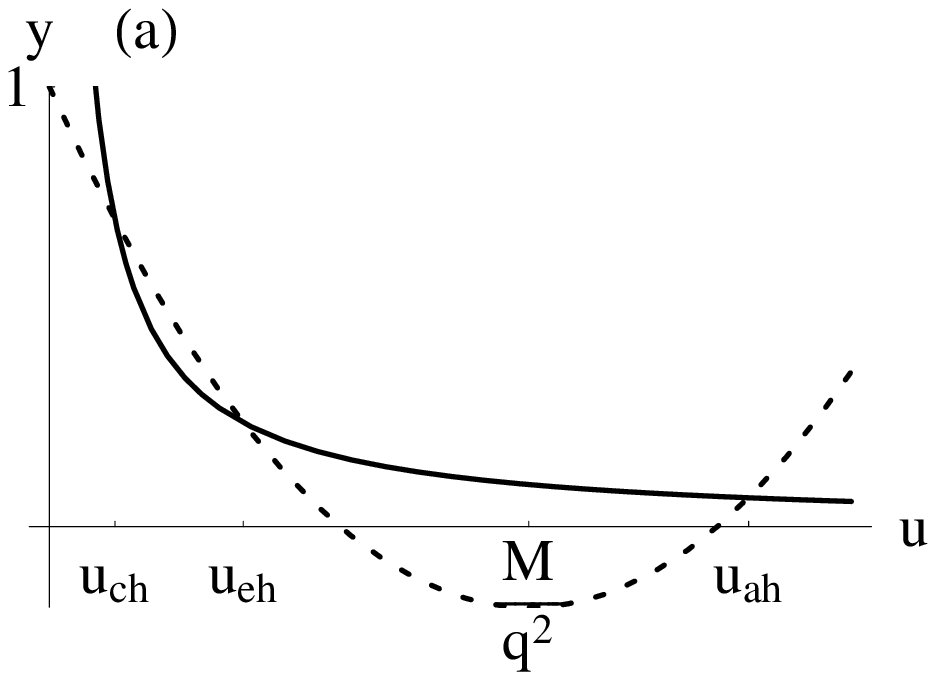} \includegraphics[width=0.32\textwidth]{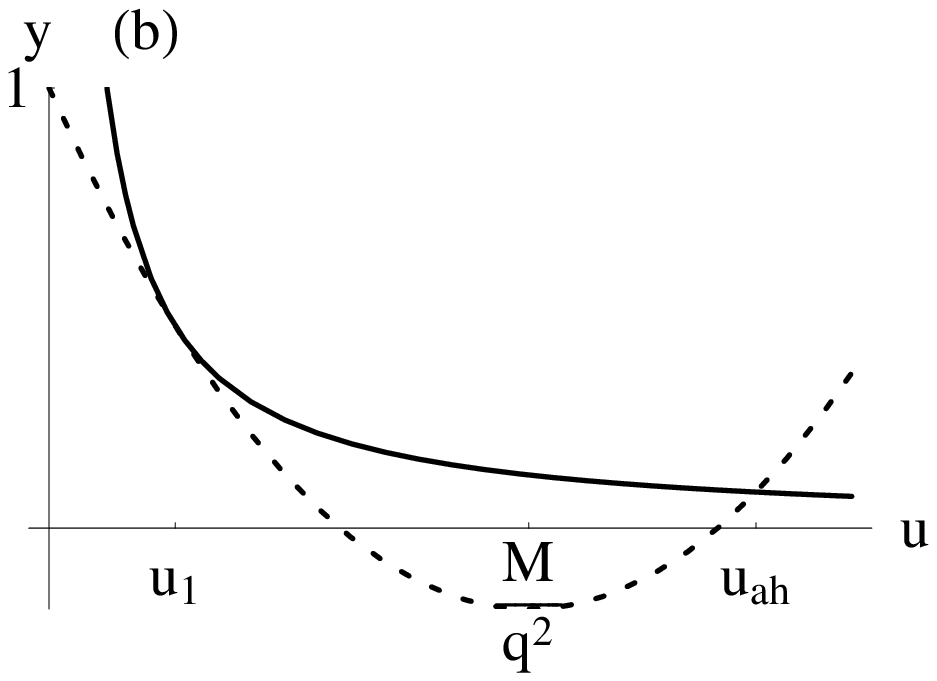} \includegraphics[width=0.32\textwidth]{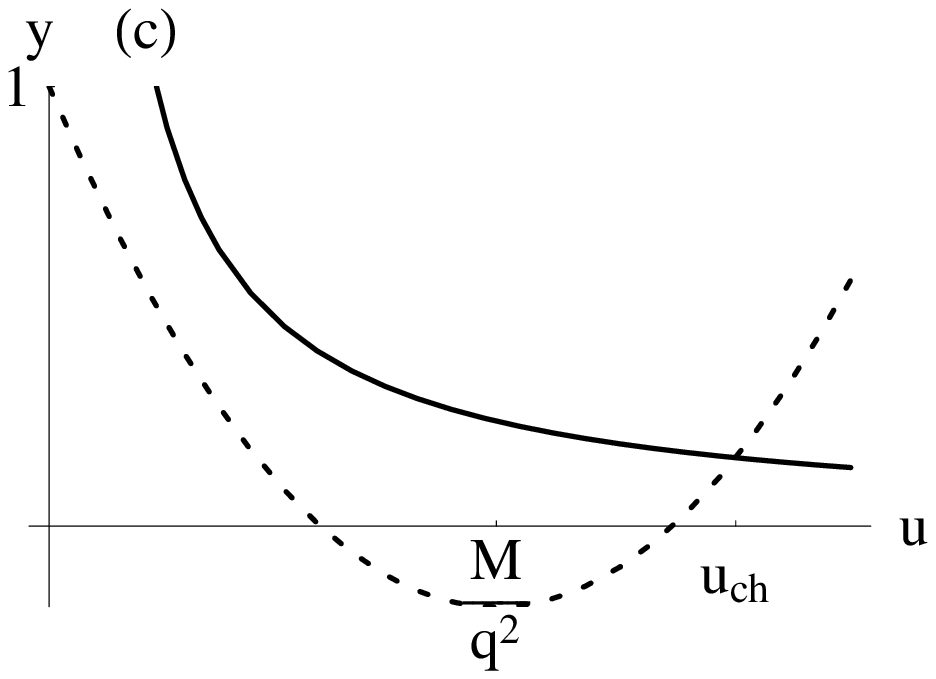}\\
  \caption{\footnotesize{Plots of $y=1-2Mu+q^2 u^2$ (dashed line) and $y=2cu^{3\om+1}$ (continuous line) for $q^2\leq M^2$ and $-2\leq 3\om+1 < 0$. (a) $c<c_1$ [Eq.~\eqref{2.5}]. (b) $c=c_1$. Here $u_1=\ch=\eh$ [Eq.~\eqref{2.6}]. (c) $c>c_1$.}}\label{Fig1}
\end{figure*}

\begin{figure*}
\centering
  \includegraphics[width=0.37\textwidth]{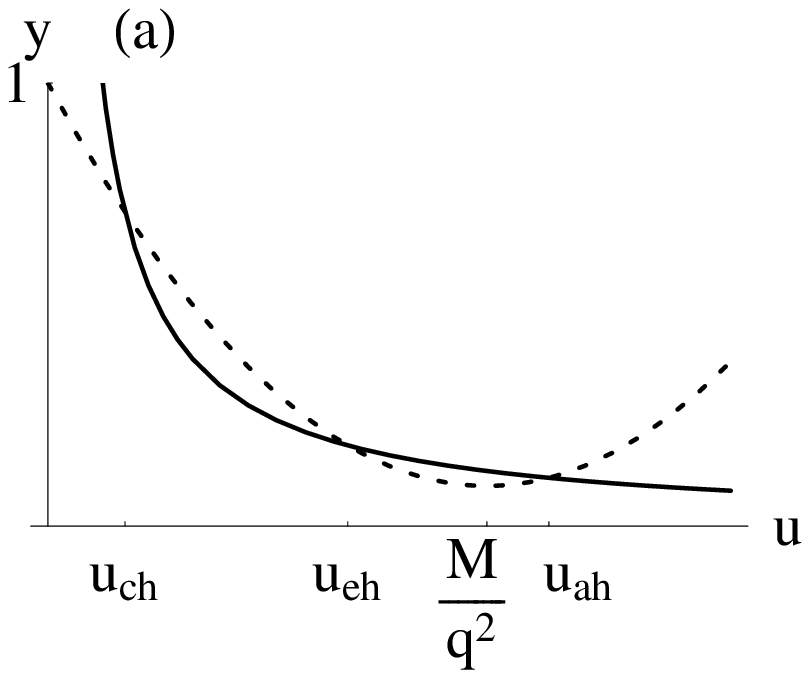} \includegraphics[width=0.37\textwidth]{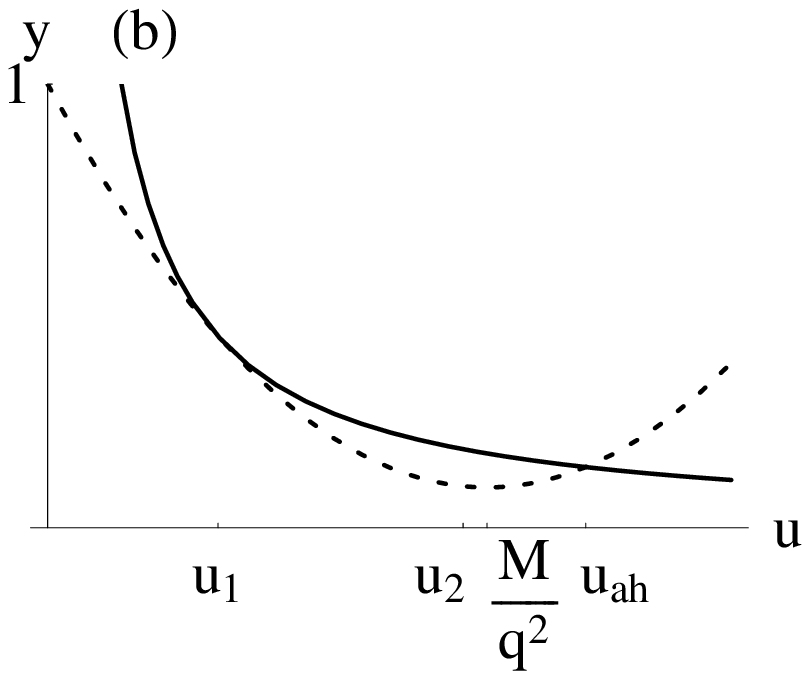} \includegraphics[width=0.37\textwidth]{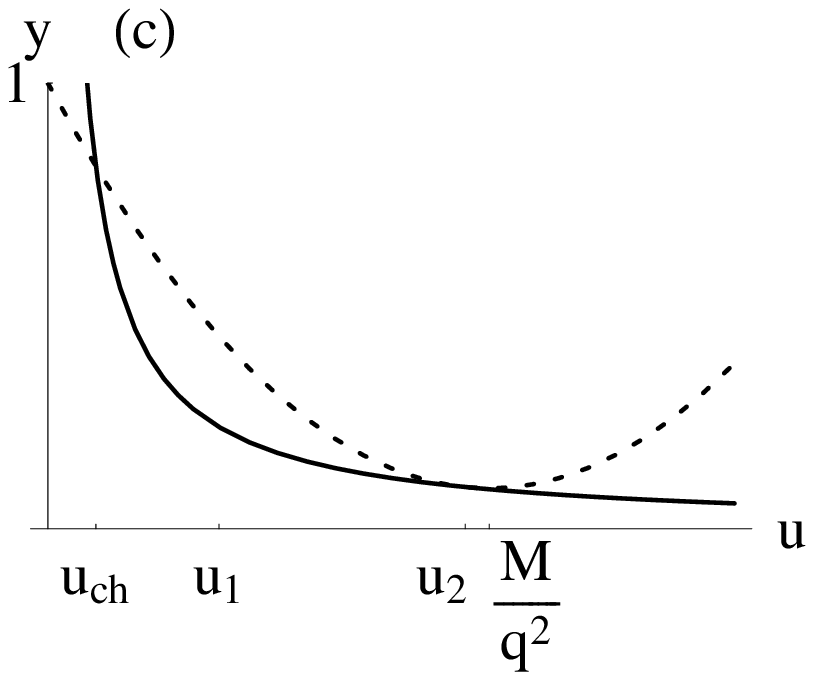} \includegraphics[width=0.37\textwidth]{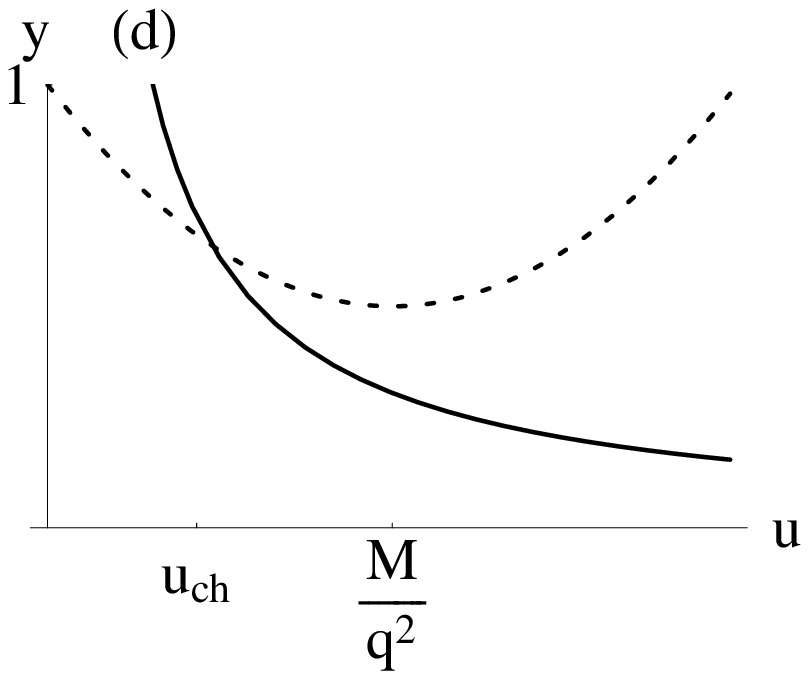}\\
  \caption{\footnotesize{Plots of $y=1-2Mu+q^2 u^2$ (dashed line) and $y=2cu^{3\om+1}$ (continuous line) for $q^2> M^2$ and $-2\leq 3\om+1 < 0$. (a) $c_2<c<c_1$ and $9\om^2M^2/(9\om^2-1)>q^2>M^2$ [Eq.~\eqref{2.7}]. (b) $c=c_1$ and $9\om^2M^2/(9\om^2-1)>q^2>M^2$. Here $u_1=\ch=\eh$ [Eq.~\eqref{2.6}]. (c) $c=c_2$ [Eq.~\eqref{2.8}] and $9\om^2M^2/(9\om^2-1)>q^2>M^2$. Here $u_2=\eh=\ah$ [Eq.~\eqref{2.9}]. (d) $q^2>9\om^2M^2/(9\om^2-1)>M^2$.}}\label{Fig2}
\end{figure*}

\begin{figure}
\centering
  \includegraphics[width=0.37\textwidth]{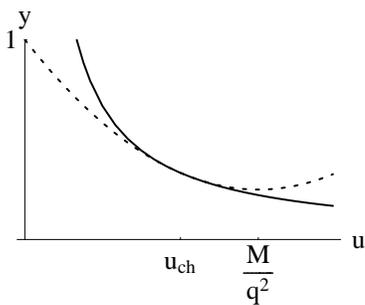}\\
  \caption{\footnotesize{Plots of $y=1-2Mu+q^2 u^2$ (dashed line) and $y=2cu^{3\om+1}$ (continuous line) for $q^2=9\om^2M^2/(9\om^2-1)>M^2$ and $-2\leq 3\om+1 < 0$. In this case, $u_1=u_2$ [Eqs.~\eqref{2.6} and~\eqref{2.9}] yielding $c_1=c_2$ [Eqs.~\eqref{2.5} and~\eqref{2.8}]. The three horizons merge.}}\label{Fig3}
\end{figure}

As is well known, the thermodynamics of singular horizons is a subtle issue. In the case of metric~\eqref{2.1}, we assert on exploring its physical properties that all the scalar invariants diverge only at the singularity $r=0$, as the density of energy and isotropic pressure do [Eq.~\eqref{2.3}]. Particularly the curvature scalar takes the form
\begin{equation}
 R=6 c \omega  (1-3 \omega ) r^{-3 (1+\omega )}.
\end{equation}
This implies that all the horizons $r_h>0$, which are solutions to $f(r)=0$, are regular. This point is important for the thermodynamic treatment we will present in Sec's.~\ref{sec4}, \ref{sec5}, and~\ref{sec5b} where no singular horizon is present.

\section{Nariai type horizons -- Extreme \bh  inside cosmological horizons -- Extreme cosmo-black-holes \label{sec3}}

From now on we restrict ourselves to asymptotically de Sitter solutions where $3\om+1 < 0$. For fixed ($M^2,q^2,\om$), the number and nature of the horizons depend on the \q charge $c$. Setting $u=1/r$, $f(r)=0$ implies
\begin{equation}\label{2.4}
   1-2Mu+q^2 u^2=2cu^{3\om+1}\,.
\end{equation}
Figures~\ref{Fig1} and~\ref{Fig2} show plots of the parabola $y=1-2Mu+q^2 u^2$ and the curve $y=2cu^{3\om+1}$ for $q^2\leq M^2$ and $q^2> M^2$, respectively. We consider these cases separately and we determine, fixing ($M^2,q^2,\om$), the constraint(s) on $c$ for which two or three horizons merge, the corresponding values of all the horizons of the solution~\eqref{2.1}, and the metric $f$.

\subsection{Case: $q^2\leq M^2$} In plot (a) of figure~\ref{Fig1} the solution~\eqref{2.1} has three horizons, a cosmological horizon ($\ch <M/q^2$ with $M/q^2$ being the value of $u$ where the parabola has a minimum), an event horizon ($\eh <M/q^2$), and an inner horizon ($\ah >M/q^2$).

In plot (b) of figure~\ref{Fig1}, the two curves have a common tangent at $u_1<M/q^2$ where $\ch$ and $\eh$ merge: this is a generalization of Nariai horizon~\cite{Nariai1,Nariai2} to all asymptotically de Sitter solutions $-1\leq\om <-1/3$. These solutions~\eqref{2.1} possess another horizon $\ah$. This happens when $c=c_1$ (see similar discussion following Eqs.~(2.11) and~(2.12) of Ref.~\cite{AAR}) with
\begin{align}
\label{2.5}&c_1=\frac{q^2u_1-M}{(3\om+1)u_1{}^{3\om}}>0 \\
\label{2.6}&u_1=-\frac{\sqrt{9\om^2M^2+(1-9\om^2)q^2}+3\om M}{(1-3\om)q^2}>0
\end{align}
and
\begin{equation*}
    \ch = \eh =u_1.
\end{equation*}

In the following we apply Eqs~\eqref{2.5} and~\eqref{2.6} to the cosmological constant case $\om =-1$ and the case $\om =-2/3$.

\subsubsection{The cosmological constant case $\om =-1$} In this case the metric~\eqref{2.1} reads
\begin{align}
&f=q^2 (u-u_1)^2 (u-\ah) (u-u_n)/u^2\nn\\
&4q^2u_1=3M-m_2,\quad (m_2\equiv \sqrt{9 M^2-8 q^2}>0)\nn\\
\label{2.6b}&4q^2\ah =M+m_2+2 \sqrt{M (M+m_2)}\\
&2q^2(u_n+\ah)=M+m_2,\quad (u_n<0)\nn\\
&\ro_{\text{q}}=\La =6c_1=3(M+m_2)(3M-m_2)^3/(4^4q^6)\nn.
\end{align}
Here $u_1$ is the common value of the horizons $\ch$ and $\eh$. In the case $q^2\leq M^2<(9/8)M^2$ of this section, $3M>m_2>0$ and it is straightforward to show that $u_n<0$. Thus, there are only three positive roots to $f=0$: $\ah>\ch=\eh>0$.

The Nariai type solution~\eqref{2.6b} generalizes the known neutral solution~\cite{Nariai1,Nariai2} to charged one. This is shown as follows. In the limit $q\to 0$, we have $\lim_{q\to0}\La =1/(9M^2)$, $\lim_{q\to0}(1/u_1)=3M$, and the inner horizon disappears in the limit $q\to 0$ as expected~\cite{Nariai1,Nariai2}. We also find: $\lim_{q\to0}(1/u_n)=-6M$ and $\lim_{q\to0}q^2\ah=2M$ yielding
\begin{multline}
\lim_{q\to 0}f=[(u-\lim_{q\to 0}u_1)^2]\Big[\lim_{q\to 0}q^2-\frac{\lim_{q\to 0}q^2\ah}{u}\Big]\times\\
\frac{(u-\lim_{q\to 0}u_n)}{u}=\\-\frac{(1-3Mu)^2(1+6Mu)}{27M^2u}
=-\frac{(r-3M)^2(r+6M)}{27M^2r}
\end{multline}
as in~\cite{Nariai1,Nariai2}.

\subsubsection{The case $\om =-2/3$} This case yields another new charged Nariai type solution:
\begin{align}
&f=q^2 (u-u_1)^2 (u-\ah)/u\nn\\
&3q^2u_1=2M-m_1,\quad (m_1\equiv \sqrt{4 M^2-3 q^2}>0)\nn\\
\label{2.6c}&3q^2\ah =2(M+m_1)\\
&c_1=(M+m_1)(2M-m_1)^2/(27q^4).\nn
\end{align}
Here again $u_1>0$ is the common value of the horizons $\ch$ and $\eh$. In the limit $q\to 0$, we have
$\lim_{q\to0}c_1 =1/(16M)$, $\lim_{q\to0}(1/u_1)=4M$, $\lim_{q\to0}q^2\ah=2M$ ($\ah$ disappears in the limit $q\to 0$), and
\begin{equation}
    \lim_{q\to0}f=-\frac{(1-4Mu)^2}{8Mu}=-\frac{(r-4M)^2}{8Mr}.
\end{equation}

The plot (a) of figure~\ref{Fig1} corresponds to $c<c_1$ and the plot (c) of the same figure, where the solution~\eqref{2.1} has only a cosmological horizon $\ch >M/q^2$, corresponds to $c>c_1$.

\subsection{Case: $q^2> M^2$} The two curves will have two common tangents (for two different values of $c$), as shown in plots (b) and (c) of figure~\ref{Fig2}, provided
\begin{equation}\label{2.7}
    1+\frac{1}{9\om^2-1}>\frac{q^2}{M^2}>1
\end{equation}
(where the non-asymptotic flat condition $-1\leq \om <-1/3$ implies $0<9\om^2-1\leq 8$).

Constraints~\eqref{2.7} being satisfied, the common tangents occur at
\begin{description}
  \item [(a)] $u_1=\ch=\eh<M/q^2$ if $c=c_1$. In the cases $\om =-1$ and $\om =-2/3$, the new charged Nariai type solutions are still given by~\eqref{2.6b} and~\eqref{2.6c}. Since the leftmost hand side of~\eqref{2.7} is 9/8 and 4/3 for $\om =-1$ and $\om =-2/3$, respectively, these two new charged Nariai type solutions generalize the previous cases~\eqref{2.6b} and~\eqref{2.6c} to $(9/8)M^2>q^2>M^2$ and $(4/3)M^2>q^2>M^2$, respectively. [plot (b) of figure~\ref{Fig2}];
  \item [(b)] $u_2=\eh=\ah<M/q^2$ if $c=c_2$ where the event horizon merges with the inner horizon yielding an extreme \abh  inside a cosmological horizon [plot (c) of figure~\ref{Fig2}]. $c_2$ and $u_2$ are given by
\begin{align}
\label{2.8}&0<c_2=\frac{q^2u_2-M}{(3\om+1)u_2{}^{3\om}}<c_1 \\
\label{2.9}&u_2=\frac{\sqrt{9\om^2M^2+(1-9\om^2)q^2}-3\om M}{(1-3\om)q^2}>u_1.
\end{align}
For this case \textbf{b)}, we again consider separately the cosmological constant case $\om =-1$ and the case $\om =-2/3$.
\end{description}

\subsubsection{The cosmological constant case $\om =-1$} The solution reads
\begin{align}
&f=q^2 (u-u_2)^2 (u-\ch) (u-u'_n)/u^2\nn\\
&4q^2u_2=3M+m_2,\quad [M>m_2 \text{ if }q^2 >M^2]\nn\\
\label{2.6d}&4q^2\ch =M-m_2+2 \sqrt{M (M-m_2)}\\
&2q^2(u'_n+\ch)=M-m_2,\quad [u'_n<0 \text{ if }q^2 >M^2]\nn\\
&\ro_{\text{q}}=\La =6c_2=3(M-m_2)(3M+m_2)^3/(4^4q^6)\nn.
\end{align}
This is an extreme \abh  inside a cosmological horizon where $u_2$ is the common value of $\eh$ and $\ah$.
In the limit $q\to M^+$ we have $\lim_{q\to M^+}c_2 =0$, $\lim_{q\to M^+}(1/u_2)=M$, and the cosmological horizon disappears ($\lim_{q\to M^+}\ch=\lim_{q\to M^+}u'_n=0$): this is the extreme \RN \aBH, as expected, where $f=(r-M)^2/r^2$.

\subsubsection{The case $\om =-2/3$} This is another extreme \abh  inside a cosmological horizon:
\begin{align}
&f=q^2 (u-u_2)^2 (u-\ch)/u\nn\\
&3q^2u_2=2M+m_1,\quad [M>m_1 \text{ if }q^2 >M^2]\nn\\
\label{2.6e}&3q^2\ch =2(M-m_1)\\
&c_2=(M-m_1)(2M+m_1)^2/(27q^4)\nn
\end{align}
In the limit $q\to M^+$ we have $\lim_{q\to M^+}c_2 =0$, $\lim_{q\to M^+}(1/u_2)=M$, and the cosmological horizon disappears ($\lim_{q\to M^+}\ch=0$): this is again the extreme \RN \abh where $f=(r-M)^2/r^2$.

If the constraints~\eqref{2.7} are still satisfied but $c_2<c<c_1$, the three horizons exist as in plot (a) of figure~\ref{Fig2}. Otherwise, if $q^2/ M^2\geq 1+1/(9\om^2-1)$, only the cosmological horizon survives as in the plot (d) of figure~\ref{Fig2}. However, when the equality holds, $q^2/ M^2= 1+1/(9\om^2-1)$, the three horizons merge, as in figure~\ref{Fig3}, and $c_1=c_2$. This case deserves a special treatment.

\subsection{Case $q^2=9\om^2M^2/(9\om^2-1)>M^2$ -- Extreme cosmo-black-holes} For this case we have simple expressions for $\ch$ and $c_1$ derived as follows. By Eqs.~\eqref{2.5}, \eqref{2.6}, \eqref{2.8} and~\eqref{2.9} we have $u_1=u_2\equiv u_H=3\om M/[(3\om -1)q^2]<M/q^2$ and $c_1=c_2\equiv c_H$ yielding
\begin{multline}\label{2.10}
    u_H =\frac{3\om +1}{3\om M},\;c_H=\frac{1}{3\om(3\om -1)}\bigg(\frac{3\om M}{3\om +1}\bigg)^{3\om +1},\\
q^2=\frac{9\om^2M^2}{9\om^2-1}.\qquad\qquad\qquad\qquad
\end{multline}
Note that for the asymptotically de Sitter solutions ($-1\leq \om <-1/3$) all factors $\om$, $3\om +1$, and $3\om -1$ in~\eqref{2.10} are negative, resulting in positive factors $\om/(3\om +1)$ and $\om(3\om -1)$. We term this type of solutions where the three horizons merge by the extreme cosmo-black-holes.

Eliminating $M$ in~\eqref{2.10}, we write the radius $r_H=1/u_H$ of the extreme cosmo-black-hole as
\begin{equation}\label{2.10a}
\hspace{-1.5mm}r_H\hspace{-0.3mm}=\hspace{-0.3mm}[3\om (3\om -1)c_H]^{\tfrac{1}{3\om +1}}\hspace{-0.3mm}=\hspace{-0.3mm}\frac{3\om M}{3\om +1}\hspace{-0.3mm}=\hspace{-0.3mm}\Big[\frac{3\om-1}{3\om+1}\Big]^{\tfrac{1}{2}}|q|.
\end{equation}
This is the most general relation expressing $r_H$ in terms of ($c_H,\om$), ($M,\om$), or ($|q|,\om$) for asymptotically de Sitter solutions.

For $\om$ held constant, $r_H$ appears as increasing linear function of $M$ and of $|q|$ with slopes $3\om/(3\om +1)$ and $[(3\om -1)/(3\om +1)]^{1/2}$, respectively. The slopes themselves are increasing functions of $\om$ varying from $3/2$ and $\sqrt{2}$, respectively, at $\om=-1$ to $\infty$ as $\om\to(-1/3)^-$.

For $M$ held constant, $r_H$ takes its minimum value $3M/2$ at $\om=-1$ and increases monotonically to indefinitely large values as $\om$ approaches $-1/3$ from the left. However, for fixed $M$, $c_H$ does not always vary monotonically. For instance, for $M=1$, $c_H$ increases monotonically to $(1/2)^-$, and for $M=0.3$, $c_H$ first decreases to some minimum value then approaches $(1/2)^-$ as $\om\to (-1/3)^-$. Using~\eqref{2.10}, we obtain for $M$ held constant
\begin{equation*}
    c_H=\tfrac{1}{2}[1+3\ep\ln\ep +O(\ep)]\qquad (\ep\equiv -\om-\tfrac{1}{3}>0),
\end{equation*}
which shows that $\lim_{\om\to (-1/3)^-}c_H=(1/2)^-$ is independent of the value of $M$ (held constant) as is $\lim_{\om\to (-1/3)^-}q^2=\infty$. In the limit $\om\to (-1/3)^-$, we have thus a huge ($r_H\to\infty$) extreme \RN cosmo-\abh with a huge electric charge but finite mass surrounded by a finite \q charge $c\to(1/2)^-$.

From a physical point of view it is rather easier to figure out configurations where $q$ is held constant than configuration where the mass parameter $M$ is. It is straightforward to establish that, for $q$ held constant, $r_H$ takes its minimum value $\sqrt{2}|q|$ at $\om=-1$ and increases monotonically to indefinitely large values as $\om$ approaches $-1/3$ from the left. Similarly to the previous case, $c_H$ does vary monotonically with $\om$ and, using~\eqref{2.10}, we obtain for $q$ held constant
\begin{equation*}
    c_H=\tfrac{1}{2}[1+\tfrac{3}{2}\ep\ln\ep +O(\ep)]\qquad (\ep\equiv -\om-\tfrac{1}{3}>0).
\end{equation*}
This also shows that $\lim_{\om\to (-1/3)^-}c_H=(1/2)^-$ is independent of the value of $q$ (held constant) as is $\lim_{\om\to (-1/3)^-}M=0$. In the limit $\om\to (-1/3)^-$, we have thus a huge ($r_H\to\infty$) but massless extreme \RN cosmo-\abh with a finite electric charge surrounded by a finite \q charge $c\to(1/2)^-$.

Finally, let us discuss the case where $M$ is taken proportional to $3\om+1$: $M=-\al (3\om+1)$ and $\al>0$. Eq.~\eqref{2.10} leads to $q^2\propto (3\om+1)$ and\footnote{In this case $c_H$ behaves as: $c_H=\tfrac{1}{2}-\tfrac{3}{4}(3+2\ln\al)\ep+\tfrac{9}{8}(3+6\ln\al+2\ln^2\al)\ep^2+O(\ep^3)$ and $\ep\equiv -\om-\tfrac{1}{3}>0$.}
\begin{align*}
&\lim_{\om\to (-1/3)^-}c_H=(1/2)^-\text{ if }\al \geq \e^{-3/2};\\
&\lim_{\om\to (-1/3)^-}c_H=(1/2)^+\text{ if }\al < \e^{-3/2}.
\end{align*}
In the limit $\om\to(-1/3)^-$, $M\to 0$, $q\to 0$, and $r_H\to\al$. There remains a pure \q state with a finite cosmological horizon $r_H\to\al$ and a finite \q charge $c\to 1/2$.

Now, we consider the special cases $\om =-1$ and $\om =-2/3$. For the cosmological constant case $\om =-1$, on applying Eqs.~\eqref{2.10} and~\eqref{2.10a}, we obtain the extreme cosmo-black-hole
\begin{multline}\label{2.11a}
\La=\frac{2}{9M^2},\,r_H=\frac{1}{\sqrt{2\La}}=\frac{3M}{2}=\sqrt{2}|q|,\\f=\frac{(2 r-3 M)^3 (2 r+9 M)}{216 M^2 r^4},\,q^2=\frac{9M^2}{8}.
\end{multline}
The case $\om =-2/3$ yields another simple extreme cosmo-black-hole
\begin{multline}\label{2.11b}
c_H=\frac{1}{12M},\,r_H=\frac{1}{6c_H}=2M=\sqrt{3}|q|,\\f=\frac{(r-2 M)^3}{6 M r^3},\,q^2=\frac{4M^2}{3}.
\end{multline}

The case $\om =-2/3$ was treated in details in~\cite{cold} where more or less equivalent formulas to~\eqref{2.6c} and~\eqref{2.6e} were given, but no general formulas as~\eqref{2.5}, \eqref{2.6}, \eqref{2.8}, and~\eqref{2.9}, which are valid for the whole range of $\om$, were derived. Similarly, the general formulas~\eqref{2.10} and~\eqref{2.10a} were not derived in~\cite{cold} but only the relation $r_H=1/(6c_H)$ was given (Eq.~(30) of~\cite{cold}) along with expressions of $M$ and $q$ in terms of $c$. In our notation~\cite{AAR}, $c$ is half its value in~\cite{cold} and half the opposite of its original value~\cite{Kis}.

\section{Conserved charges and thermodynamics \label{sec4}}

In this work, we only consider the thermodynamics of the \textit{event horizon} and, from now on, we restrict ourselves to non-extremal solutions, that is, to cases where the three horizons do not merge with each other. For the asymptotically de Sitter solutions ($-1\leq \om <-1/3$), these are the solutions satisfying either one of the two following constraints:
\begin{align}
\label{4.4a}&q^2\leq M^2 \text{ and } c<c_1,\text{ or},\\
\label{4.4b}&\frac{9\om^2 M^2}{9\om^2-1}>q^2>M^2 \text{ and } c_2<c<c_1.
\end{align}
Solutions satisfying the first constraint correspond to plot (a) of figure~\ref{Fig1} and those satisfying the second constraint correspond to plot (a) of figure~\ref{Fig2}.

The temperature of the event horizon is given by
\begin{multline}\label{4.5}
\teh = \frac{\partial_r f}{4\pi}\Big|_{\reh}\\=\frac{\eh}{2\pi}[M-q^2\eh+(3\om+1)c\eh{}^{3\om}],
\end{multline}
where one may eliminate $M$ using~\eqref{2.4}. Note that $-2M+2q^2\eh$ and $2(3\om+1)c\eh{}^{3\om}$ are the derivatives, evaluated at the point $\eh$, of the functions $y=1-2Mu+q^2 u^2$ and $y=2cu^{3\om+1}$, respectively, which are plotted in figures~\ref{Fig1} and~\ref{Fig2}. From the plots (a) of these two figures one sees that the slopes are such that
\begin{equation}\label{4.6}
    2(3\om+1)c\eh{}^{3\om}>-2M+2q^2\eh,
\end{equation}
which yields $\teh>0$ for the \bh constrained by~\eqref{4.4a} or~\eqref{4.4b}.

As far as $\teh>0$, the evaporation of the \abh proceeds by reducing the mass parameter $M$. Differentiating~\eqref{2.4} with respect to $\eh$ yields
\begin{equation}\label{4.7}
   -\Big(\frac{\partial M}{\partial u}\Big)\Big|_{\eh}= \frac{M-q^2\eh+(3\om+1)c\eh{}^{3\om}}{\eh}\propto \teh,
\end{equation}
or, equivalently,
\begin{equation}\label{4.8}
   \Big(\frac{\partial M}{\partial r}\Big)\Big|_{\reh}\propto \teh >0.
\end{equation}
Using the plots (a) of figures~\ref{Fig1} and~\ref{Fig2} one obtains at $\ch$ and $\ah$ similar inequalities to~\eqref{4.6} but with the other order sign: $2(3\om+1)c\ch{}^{3\om}<-2M+2q^2\ch$ and $2(3\om+1)c\ah{}^{3\om}<-2M+2q^2\ah$. This yields
\begin{equation}\label{4.10}
   \Big(\frac{\partial M}{\partial r}\Big)\Big|_{\rch}<0\;\text{ and }\;\Big(\frac{\partial M}{\partial r}\Big)\Big|_{\rah}<0,
\end{equation}
where we have used the fact that $\partial M/\partial r\propto M-q^2u+(3\om+1)cu^{3\om}$.

Using~\eqref{4.8} and~\eqref{4.10} we conclude that during the evaporation, the event horizon shrinks and the other two horizons expand. For \bh constrained by~\eqref{4.4a} [respectively by~\eqref{4.4b}], as the evaporation proceeds the configuration evolves from the plot (a) of figure~\ref{Fig1} [respectively from the plot (a) of figure~\ref{Fig2}] to the plot (c) of figure~\ref{Fig2} where $\teh=0$ since the two slopes are equal. It is worth mentioning that the configuration evolve directly from the plot (a) of figure~\ref{Fig1} [respectively from the plot (a) of figure~\ref{Fig2}] to the plot (c) of figure~\ref{Fig2} without passing by the other configurations shown in figures~\ref{Fig1} and~\ref{Fig2}.

During the evaporation, all the other given parameters ($q,c,\om$) are held constant but the mass parameter, which decreases from some initial value $M$, constrained by~\eqref{4.4a} or~\eqref{4.4b}, to some final value $M_f$ satisfying
\begin{equation}\label{4.n1}
    (9\om^2-1)q^2/9\om^2<M_f{}^2<q^2.
\end{equation}
The evaporation ends when the value of $c_2$, which depends on $M$ as given in~\eqref{2.8}, ceases to vary and settles to the given value of $c$. The final mass of the extreme \abh inside a cosmological horizon is solution to the extremality condition [compare with~\eqref{2.8} and~\eqref{2.9}]
\begin{equation}\label{4.n2}
  \frac{q^2u_2-M_f}{(3\om+1)u_2{}^{3\om}}=c,
\end{equation}
which expresses the equality of the slopes in the plot (c) of figure~\ref{Fig2}, and $u_2$ is given by
\begin{equation}\label{4.n3}
    u_2=\frac{\sqrt{9\om^2M_f{}^2+(1-9\om^2)q^2}-3\om M_f}{(1-3\om)q^2}.
\end{equation}

Equations~\eqref{4.n2} and~\eqref{4.n3} reproduce the correct result for the extreme \RN \aBH. In absence of \Q, $c=0$ and Eq.~\eqref{4.n2} gives $u_2=M_f/q^2$. In order to derive from Eq.~\eqref{4.n3} this same value for $u_2$, which does not depend on $\om$, the only possibility is to set there $q^2=M_f{}^2$ which results in $u_2=M_f/q^2$.

As we shall see below, the static \dS spacetimes have different thermodynamic notions of energy but the above results derived in this section are independent of these different notions.

Rotating or static \abh solutions, of general relativity or extended theories, immersed in flat or anti-de Sitter spacetimes have well defined physical entities, these are the globally so-called \textit{charges} (mass, electric and magnetic charges, angular momentum, scalar charges and so on). These solutions may have multi-horizons among which one finds only one event horizon but no cosmological horizon.

By the static \dS spacetimes we mean all solutions where one of the metric components is of the form $1-2M/r+q^2/r^2-\La r^2/3$, with $\La>0$ and $M$ and $q$ are constants [in the charged Vaidya-\dS \aBH, $M$ and $q$ are not constants but the metric is nonstatic~\cite{V1,V2}]. The static \dS spacetimes, with a non-extremal event horizon and a cosmological horizon, have the property that the thermodynamic temperatures of these two horizons are not equal~\cite{GH}.

We enlarge the above list of static \dS spacetimes by including all \bh with a non-extremal event horizon and a cosmological horizon, as the solutions given by~\eqref{2.1}, \eqref{2.2}, and~\eqref{2.3b}. This is called the set of the de Sitter-like spacetimes\footnote{There are other static \BH, with a non-extremal event horizon and a cosmological horizon, but where the mass parameter depends on $r$: $M\equiv M(r)$~\cite{irina}. These make part of the de Sitter-like spacetimes but we will not include them in our discussion.}. They have the property that, in general, the thermodynamic temperatures of the event and cosmological horizons are not equal (this property is violated by some black holes with conformally coupled scalar field, as the  Mart\'{\i}nez-Troncoso-Zanelli ones~\cite{MTZ}, where the two horizons have the same temperature~\cite{C5}).

In the Euclidean formulation, this means that the imaginary time periods are not equal and, consequently, it is not possible to avoid the conical singularities at both horizons at once. Said otherwise, if one of the two horizons is treated as a thermodynamic system, the other horizon cannot be treated so because of the presence of the conical singularity there; it is, rather, treated as a boundary. Considerations by which both horizons are treated simultaneously as different thermodynamic systems, out of equilibrium, are subtle. An instance of that, one cannot add the entropies of the two horizons to obtain the total entropy of the ``thermodynamic system made out of the two horizons", which, in fact, it does not exist.

The other issue with the de Sitter-like spacetimes is the definition and evaluation of the conserved quantities. Because of no spatial infinity accessible to observers, the notions of ADM mass, electric charge and other charges, which are defined in asymptotically flat or anti-de Sitter spacetimes, are no longer valid. Different prescriptions to define conserved charges exist however. These are known as Abbott-Deser (AD)~\cite{AD}, Balasubramanian-Boer-Minic (BBM)~\cite{BBM}, and Teitelboim~\cite{Tt1,Tt2} prescriptions. A statement made in Ref.~\cite{Sekiwa} asserts that the three prescriptions yield the same conserved charges when applied to static asymptotically de Sitter solutions: ``Furthermore, for nonrotating case Teitelboim's charges are in full agreement with BBM/AD charges".

For the static de Sitter-like spacetimes, it is straightforward to generalize the expression of the Teitelboim's energy to these holes provided (1) the mass parameter $M$ is constant [see Eq.~(3.19) of~\cite{Tt1} and see Eq.~(14) of~\cite{Tt2}] and (2) that no other parameter in the metric depends on $M$ [for the case of the solutions given by~\eqref{2.1}, \eqref{2.2}, and~\eqref{2.3b}, the constants ($q,c$) must not depend on $M$ in order to generalize the Teitelboim's energy expression]. If $E_{\text{T}}$ denotes the Teitelboim's energy of the \textit{event horizon}, then
\begin{equation}\label{4.1}
    E_{\text{T}}=M.
\end{equation}

In classical thermodynamics, the thermodynamic description of a system could be achieved using different forms of energy, which yields the notions of ensembles: micro-canonical, canonical, and grand-canonical. For the \dS spacetimes, it seems that the notion of ensembles is larger than what one usually encounters in classical thermodynamics. This is because there is not a universally agreed definition of the energy for this class of \bh with two or more horizons. Even within some elected definition of the energy, say Teitelboim's definition~\eqref{4.1}, it is possible to have some new emerged ensembles.

Another perception of the notion of energy for the \dS spacetimes is due to Padmanabhan~\cite{Pad1,Pad2,Pad3}. Using a path integral approach one derives the Padmanabhan's energy $E_{\text{P}}$ of the \textit{event horizon} by
\begin{equation}\label{4.2}
    E_{\text{P}}=\reh/2=1/(2\eh),
\end{equation}
where $\eh$ is a solution to~\eqref{2.4}. Using the latter equation we relate the two energies by
\begin{equation}\label{4.3}
   E_{\text{T}}=E_{\text{P}}+\frac{q^2}{4}\,\frac{1}{E_{\text{P}}}-\frac{c}{2^{3\om}}\,\frac{1}{E_{\text{P}}{}^{3\om}}.
\end{equation}

The cosmological horizon has its corresponding, and similar, formulas to~\eqref{4.1} and~\eqref{4.2}.

Since the evaporation of the \abh proceeds by reducing the mass parameter $M$. By~\eqref{4.1}, this results in a reduction of $E_{\text{T}}$ and should yield the same for $E_{\text{P}}$. This is in fact the case since Eq.~\eqref{4.8} is just
\begin{equation}\label{4.9}
   \frac{\partial E_{\text{T}}}{\partial E_{\text{P}}} >0.
\end{equation}
Hence, $\dd E_{\text{T}}$ and $\dd E_{\text{P}}$ have the same signs for the \bh constrained by~\eqref{4.4a} or~\eqref{4.4b}.

The evaporation ends when $\teh=0$ and $E_{\text{T}}$ reaches its minimal value $M_f$ given by Eqs.~\eqref{4.n2} and~\eqref{4.n3}. $E_{\text{P}}$ also reaches its minimal value $E_{\text{P},f}$ at the end of the process where $E_{\text{P},f}=1/(2u_2)$ with $u_2$ given by~\eqref{4.n2} and~\eqref{4.n3}.

As claimed earlier in this section, for the \dS spacetimes, the notion of ensembles is larger than what one usually encounters in classical thermodynamics. This is why in this paper, we will not employ the classical thermodynamic terminology of micro-canonical, canonical, and grand-canonical; rather, we will describe ensembles by the constancy of the corresponding thermodynamic variable or potential. Instances are provided by the ensembles where either $E_{\text{T}}$ or $E_{\text{P}}$ is held constant. The ensembles $E_{\text{T}}=C_1$ and $E_{\text{P}}=C_2$, where ($C_1,C_2$) are constants, describe different thermodynamic systems since $E_{\text{T}}$ held constant is represented in the 3-dimensional space ($E_{\text{P}},q,c$) by the 2-dimensional curved surface~\eqref{4.3} where $E_{\text{T}}=C_1$. This shows that there is, in fact, no ambiguity in the definition of the energy for the \dS spacetimes: $E_{\text{T}}$ and $E_{\text{P}}$ are different energies just because they correspond to different ensembles, exactly in the same way as the Gibbs free energy differs from the Helmholtz free energy.

\section{Event horizon thermodynamics: Enthalpy versus internal energy \label{sec5}}

Since the Teitelboim's energy $E_{\text{T}}=M$, we will, for time to time, insert $E_{\text{T}}$ in a couple of mathematical expressions of this section to show its relation to some thermodynamic potentials.

With the entropy of the event horizon given by
\begin{equation}\label{4a1}
    S=\pi \reh{}^2=\pi s \qquad (s\equiv S/\pi),
\end{equation}
we re-write the expression of $M$, using $1-2M\eh+q^2 \eh{}^2=2c\eh{}^{3\om+1}$ [see Eq.~\eqref{2.4}], as
\begin{equation}\label{4a2}
M=\frac{\sqrt{s}}{2}+\frac{q^2}{2\sqrt{s}}-\frac{c}{s^{3\om/2}}.
\end{equation}

Considering ($s,q,c$) as independent thermodynamic variables and using~\eqref{4a2}, it is easy to establish\footnote{Equation~\eqref{4a3} was derived in Ref.~\cite{AAR} for asymptotically flat solutions ($-1/3\leq \om<0$), however, the derivation is purely mathematical and applies also to the case of asymptotically de Sitter solutions we are considering in this paper ($-1\leq \om <-1/3$). The derivation stems from the fact that $M$, as given by~\eqref{4a2}, is homogeneous in ($S,q^2,c^{2/(3\om+1)}$) of order $1/2$.} the generalized Smarr formula~\cite{AAR}
\begin{equation}\label{4a3}
    M=2\teh S+Aq+(3\om+1)\Ta c
\end{equation}
where
\begin{equation}\label{4a4}
    A=\Big(\frac{\partial M}{\partial q}\Big)_{S,c}=\frac{q}{\sqrt{s}},\;\Ta=\Big(\frac{\partial M}{\partial c}\Big)_{S,q}=-\frac{1}{s^{3\om/2}},
\end{equation}
are the electric potential on the event horizon and the thermodynamic conjugate of $c$, respectively.
It is straightforward to check that $\teh$ as defined in~\eqref{4.5} is just
\begin{equation}\label{4a5}
    \teh=\Big(\frac{\partial M}{\partial S}\Big)_{q,c}.
\end{equation}
The first law of thermodynamics takes then the form
\begin{equation}\label{4a6}
    \dd M=\teh \dd S+A\dd q+\Ta\dd c.
\end{equation}

The last term in~\eqref{4a6} does not have a direct physical meaning; rather, we prefer to introduce a new thermodynamic variable and its conjugate which both have a familiar physical meaning. These variables are the value of the pressure $p_{\text{q}}$ evaluated at the event horizon $P\equiv p_{\text{q}}|_{\reh}$ and its conjugate, the thermodynamic volume, $V$. Using~\eqref{2.3} along with~\eqref{4a1} we obtain
\begin{equation}\label{4a7}
  P=-\frac{3\om^2 c}{4\pi s^{3(\om+1)/2}}<0.
\end{equation}
[In the cosmological constant case $\om=-1$ and $\La=6c$, $P$ reduces to constant pressure $P_{\La}=-\La/(8\pi)$]. In terms of the new independent thermodynamic variables ($s,q,P$), Eq.~\eqref{4a2} takes the form
\begin{equation}\label{4a8}
M=\frac{\sqrt{s}}{2}+\frac{q^2}{2\sqrt{s}}+\frac{4\pi}{3}\frac{s^{3/2}}{\om^2}P
\end{equation}
where it appears as homogeneous in ($S,q^2,P^{-1}$) of order $1/2$. The Euler identity for thermodynamic potentials
that are not homogeneous functions of their natural extensive variables yields~\cite{on-gtd}
\begin{equation}\label{4a9}
  M=2\Big(\frac{\partial M}{\partial S}\Big)_{q,P}S+\Big(\frac{\partial M}{\partial q}\Big)_{S,P}q-2\Big(\frac{\partial M}{\partial P}\Big)_{S,q}P,
\end{equation}
with $(\partial M/\partial q)_{S,P}=(\partial M/\partial q)_{S,c}=A$ but $(\partial M/\partial S)_{q,P}\neq (\partial M/\partial S)_{q,c}=\teh$ if $\om\neq -1$ while $M/\partial S)_{q,P}= (\partial M/\partial S)_{q,c}=\teh$ if $\om =-1$. Thus, if $\om\neq -1$ the differential of $M$ produces something which does not look like a familiar thermodynamic first law or its equivalencies
\begin{equation}\label{4a10}
    \dd M=\Big(\frac{\partial M}{\partial S}\Big)_{q,P}\dd S+A\dd q+\Big(\frac{\partial M}{\partial P}\Big)_{S,q}\dd P.
\end{equation}
This shows that $E_{\text{T}}=M$ is not a familiar thermodynamic potential.

For $\om =-1$, Eq.~\eqref{4a10} reduces to the following familiar formula:
\begin{equation}\label{4a11}
    \dd M=\teh\dd S+A\dd q+\frac{4\pi}{3}s^{3/2}\dd P_{\La},
\end{equation}
where
\begin{equation}\label{4a11b}
M=E_{\text{T}}= \frac{\sqrt{s}}{2}+\frac{q^2}{2\sqrt{s}}+\frac{4\pi}{3}s^{3/2}P_{\La}\quad \big(P_{\La}=\frac{-\La}{8\pi}\big),
\end{equation}
is interpreted as the enthalpy and
\begin{equation}\label{4a12}
V_{\La}=4\pi s^{3/2}/3,
\end{equation}
the conjugate of $P$, is the thermodynamic and geometric volume excluded from a spatial slice by the black hole horizon. This interpretation given to the mass parameter $M$ ($=E_{\text{T}}$) of the static \dS spacetimes extends that for the static anti-\dS spacetimes~\cite{ex0}-\cite{Mann2}.

We aim to extend this interpretation to all the de Sitter-like spacetimes. For the purpose of this paper, we will do that for the de Sitter-like spacetimes given by~\eqref{2.1}, \eqref{2.2}, and~\eqref{2.3b}, that is, we will enlarge the scope of the above-made interpretation to include the cases where $\om\neq -1$ by introducing a new thermodynamic potential. The new thermodynamic potential $H$ is defined such that
\begin{align*}
&\Big(\frac{\partial H}{\partial S}\Big)_{q,P}=\teh \equiv\Big(\frac{\partial M}{\partial S}\Big)_{q,c}, \\
&\Big(\frac{\partial H}{\partial q}\Big)_{S,P}=A \equiv\Big(\frac{\partial M}{\partial q}\Big)_{S,c}.
\end{align*}
This is achieved upon adding to $M$, given by~\eqref{4a8}, the following term:
\begin{equation*}
-\frac{4\pi}{3}\frac{\om+1}{\om^2}s^{3/2}P,
\end{equation*}
which is $0$ if $\om=-1$. This yields
\begin{equation}\label{4a13}
H\equiv \frac{\sqrt{s}}{2}+\frac{q^2}{2\sqrt{s}}-\frac{4\pi}{3}\frac{s^{3/2}}{\om}P,
\end{equation}
where $P$ is given by~\eqref{4a7}. Equation~\eqref{4a13} reduces to~\eqref{4a11b} if $\om=-1$. $H$ is homogeneous in ($S,q^2,P^{-1}$) of order $1/2$. With $(\partial H/\partial S)_{q,P}=\teh$ and $(\partial H/\partial q)_{q,P}=A$, the Euler identity yields
\begin{equation}\label{4a14}
  H=2\teh S+Aq-2VP,
\end{equation}
where $V$ is the thermodynamic volume, conjugate of $P$, given by
\begin{equation}\label{4a15}
    V\equiv \Big(\frac{\partial H}{\partial P}\Big)_{S,q}=-\frac{4\pi}{3}\frac{s^{3/2}}{\om}>0,
\end{equation}
which reduces to the geometric volume~\eqref{4a12} if $\om=-1$. The differential of $H$ leads to the familiar well known first-law-equivalent formula
\begin{equation}\label{4a16}
    \dd H = \teh \dd S+A\dd q+V\dd P,
\end{equation}
by which $H$ is interpreted as the enthalpy.

We have thus reached the conclusion that the Teitelboim's energy is in general not the internal energy or enthalpy of the event horizon of the de Sitter-like spacetime. The Teitelboim's energy or the mass parameter is related to the enthalpy by
\begin{equation}\label{4a17}
M=H+\frac{4\pi}{3}\frac{\om+1}{\om^2}s^{3/2}P
\end{equation}
and the internal energy $U=H-PV$ is related to $M$ by
\begin{equation}\label{4a18}
U=M-\frac{4\pi}{3}\frac{s^{3/2}}{\om^2}P.
\end{equation}
The first law should read
\begin{equation*}
    \dd U=\teh \dd S+A\dd q-P\dd V,
\end{equation*}
but since $V$ depends on $S$ via~\eqref{4a15}, one of the two variables, $S$ or $V$, is redundant. This shows that $U$ is not a convenient thermodynamic potential for expressing the first law for the static de Sitter-like spacetimes.

With $M$ given by~\eqref{4a8}, the r.h.s of~\eqref{4a18} reduces to a function of ($S,q$) only
\begin{equation}\label{4a19}
U=\frac{\sqrt{s}}{2}+\frac{q^2}{2\sqrt{s}},
\end{equation}
as it could be reduced, using~\eqref{4a15}, to a function of ($V,q$) only.

This has been noticed for the static anti-de Sitter spacetimes where the thermodynamic volume $V$ depends on the entropy $S$ too so that ``they cannot be varied independently and so $V$ seems redundant. Indeed this may be the reason why $V$ was never considered in the early literature on black hole thermodynamics. But this is an artifact of the non-rotating
approximation, $V$ and $S$ can, and should, be considered to be independent variables for a rotating black hole"~\cite{Dolan}.

We should be able to do the same upon including rotation; we may pursue that in a subsequent work. It is worth noticing that $\om$, being dimensionless, cannot be considered as a thermodynamic variable.

It is also worth noticing that the expression of $H$, as given by~\eqref{4a13} and~\eqref{4a7}, is totally independent of the definition of Teitelboim's energy. Throughout this section we have used the mass parameter $M$ for the derivation of the expression~\eqref{4a13} of $H$. Since $E_{\text{T}}=M$, we have, from time to time, inserted $E_{\text{T}}$ to show the relation of $E_{\text{T}}$ to the enthalpy, as in~\eqref{4a11b}.

We verify that the conjecture made in Ref.~\cite{iso} concerning the Reverse Isoperimetric Inequality remains true for Reissner-Nordstr\"om black holes surrounded by quintessence. This large inequality reads
\begin{equation}\label{eq1}
    \bigg(\frac{(D-1)V}{\mathcal{A}_{D-2}}\bigg)^{\tfrac{1}{D-1}}\geq \bigg(\frac{A}{\mathcal{A}_{D-2}}\bigg)^{\tfrac{1}{D-2}},
\end{equation}
where $D$ is the dimension of the spacetime, $A$ is the area of the event horizon, and $\mathcal{A}_{D-2}$ is the area of the unit $(D - 2)$-sphere. In our case, $D=4$, $\mathcal{A}_{2}=4\pi$, $A=4\pi s$, and $V$ is given by~\eqref{4a15}. This yields
\begin{equation*}
    1\geq \sqrt[3]{-\om},
\end{equation*}
which is always true and the equality holds for Reissner-Nordstr\"om-de Sitter black hole.

\section{Reissner-Nordstr\"om-de Sitter black hole surrounded by quintessence \label{sec5b}}

Up to now, we only considered separately the case of (1) the Reissner-Nordstr\"om-de Sitter black hole, that is, the Reissner-Nordstr\"om black hole surrounded by a cosmological density, and the case of (2) the Reissner-Nordstr\"om black hole surrounded by quintessence. We aim to extend the results of Sec.~\ref{sec5} to the case of the Reissner-Nordstr\"om-de Sitter black hole surrounded by quintessence, that is, to the case where the Reissner-Nordstr\"om black hole is surrounded by a cosmological density and quintessence. On doing this we extend the phase space by including two physical constants ($\La, c$) the variations of which yields two thermodynamic volumes. This extension should apply to any fundamental theory with many physical constants~\cite{iso}.

The metric $f$ now takes the form
\begin{multline}\label{f1}
f(r)=1-\frac{2M}{r}+\frac{q^2}{r^2}-\frac{\La r^2}{3}-\frac{2c}{r^{3\om+1}},\\\text{ with } -1< \om<0,\;\La >0, \text{ and } c>0.
\end{multline}
The mass parameter $M$ is expressed in terms of $s$ by [compare with~\eqref{4a2}]
\begin{equation}\label{f2}
M=\frac{\sqrt{s}}{2}+\frac{q^2}{2\sqrt{s}}-\frac{\La s^{3/2}}{6}-\frac{c}{s^{3\om/2}}.
\end{equation}
The temperature of the event horizon is no longer given by $\teh$; rather, it is given by
\begin{equation}\label{f3}
    T=\Big(\frac{\partial M}{\partial S}\Big)_{q,\La,c}.
\end{equation}

It is straightforward to generalize the results of Sec.~\ref{sec5}. For instance, Eq.~\eqref{4a3} becomes
\begin{equation}\label{f4}
    M=2TS+Aq-2\Ta_{\La} \La+(3\om+1)\Ta c,
\end{equation}
where $\Ta\equiv (\partial M/\partial c)_{S,q,\La}$ is as given in~\eqref{4a4} and $\Ta_{\La}\equiv (\partial M/\partial \La)_{S,q,c}=-s^{3/2}/6$. In a similar way we generalize Eqs.~\eqref{4a13}, \eqref{4a14} and~\eqref{4a16} to
\begin{align}
&H= \frac{\sqrt{s}}{2}+\frac{q^2}{2\sqrt{s}}+\frac{4\pi}{3}s^{3/2}P_{\La}-\frac{4\pi}{3}\frac{s^{3/2}}{\om}P,\\
&H=2T S+Aq-2V_{\La}P_{\La}-2VP,\\
&\dd H = T\dd S+A\dd q+V_{\La}\dd P_{\La}+V\dd P,
\end{align}
where $A$, $P$, $P_{\La}$, $V_{\La}$, and $V$ have the same expressions as in Sec.~\ref{sec5}. $T$ is either given by~\eqref{f3} or by
\begin{equation}
    T=\Big(\frac{\partial H}{\partial S}\Big)_{q,P_{\La},P}.
\end{equation}

The internal energy is defined by $U=H-V_{\La}P_{\La}-VP$ and retains its expression given by~\eqref{4a19} as does the expression of $M$ given by~\eqref{4a17}.

That the internal energy has the same expression for a Reissner-Nordstr\"om \abh and a Reissner-Nordstr\"om-de Sitter \abh both surrounded by \q seems to be a universal law that applies, not only to all de Sitter-like spacetimes, but to all static charged \bh even in the case where the mass parameter depends $r$. In Ref~\cite{new}, we show that the internal energy of any static charged \aBH, with possibly a variable mass parameter, is given by
\begin{equation}\label{in}
U=\frac{\sqrt{s}}{2}+\frac{q^2}{2\sqrt{s}}.
\end{equation}
This depends only on the entropy of the event horizon and on the electric charge, which are the intrinsic properties of the \aBH, and it does not depend on any extrinsic properties, as a cosmological density, \Q, or any other force that may exert a pressure on the \aBH. We have thus the following conclusion:\\

\emph{The internal energy of any static charged \aBH, with possibly a variable mass parameter, does depend explicitly only on the intrinsic properties of the black hole.}\\

However, $U$ depends implicitly on $M$ and other physical constants through $s$ which is a solution to $f(s)=0$. It is worth noticing that the first term $\sqrt{s}/2=\reh/2$ is just Padmanabhan's energy and the second term $q^2/(2\sqrt{s})=q^2/(2\reh)$ is an electric-energy contribution.

For \Sd and \RN \BH, $U$ coincides with the mass $M$ and the pressure $P$ is identically zero.

Finally, the Reverse Isoperimetric Inequality~\eqref{eq1} is satisfied separately for $V_{\La}$ and for $V$.

\section{Conclusion \label{sec6}}

We have determined the exact general conditions under which extreme solutions exist for the Reissner-Nordstr\"om black holes surrounded by quintessence with a negative quintessencial pressure. For $q^2\leq M^2$, the only existing extreme solutions are generalizations of Nariai \BH. For $q^2> M^2$, but $q^2<9\om^2 M^2/(9\om^2-1)$, we may have both extreme solutions: extreme \bh inside cosmological horizons or generalizations of Nariai \BH.

In the limit case $q^2=9\om^2 M^2/(9\om^2-1)$ we were led to the extreme cosmo-black-hole solution where all horizons merge. The limit $\om\to(-1/3)^-$ is characterized by the presence of
\begin{description}
  \item (1) a huge extreme \RN cosmo-\abh with a huge electric charge but finite mass surrounded by a finite \q charge and vanishing event horizon pressure if the mass is held constant as the limit is approached;
  \item (2) a massless huge extreme \RN cosmo-\abh with a finite electric charge surrounded by a finite \q charge and vanishing event horizon pressure if the electric charge is held constant as the limit is approached;
  \item (3) a massless and neutral extreme \RN cosmo-\aBH, with a finite radius, surrounded by a finite \q charge and nonvanishing event horizon pressure if the mass remains proportional to $3\om+1$ as the limit is approached.
\end{description}

We have shown that during the evaporation, the event horizon shrinks and the other two horizons expand, and that the final mass at the end of the evaporation is independent of the initial order relation between the squares of the electric charge and the mass parameter, provided the three horizons exist at the beginning of the process.

The inclusion of the $P$-$V$ term has led to a consistent thermodynamic description of the first law of thermodynamics. The results obtained here generalize the results obtained for the anti-de Sitter spacetime, where the pressure exerted on the horizon is positive, as well as the results for the \dS one~\cite{ex0}-\cite{Mann2}. This shows that the sign of the pressure is irrelevant. We have commented that the internal energy has a universal expression for any static charged \aBH, with possibly a variable mass parameter. We have also shown that the Reverse Isoperimetric Inequality holds.

The results concerning the thermodynamics were easily generalized to the case of the Reissner-Nordstr\"om-de Sitter black hole surrounded by quintessence with two physical constants yielding two thermodynamic volumes.

Phase transitions and critical phenomena will be discussed elsewhere.




\begin{thebibliography}{99}

\bb{ex0}M.M.~Caldarelli, G.~Cognola, and D.~Klemm,
\emph{Thermodynamics of Kerr-Newman-AdS black holes and conformal field theories},
Class. Quantum Grav. \textbf{17} 399 (2000). arXiv:hep-th/9908022.

\bb{Wang-Wu}S.~Wang, S.Q.~Wu, F.~Xie, and L.~Dan,
\emph{The first laws of thermodynamics of the (2+1)-dimensional BTZ black holes and Kerr-de Sitter spacetimes},
Chin. Phys. Lett. \textbf{23} 1096 (2006). arXiv:hep-th/0601147.

\bb{Sekiwa}Y.~Sekiwa,
\emph{Thermodynamics of de Sitter black holes: Thermal cosmological constant},
Phys. Rev. D \textbf{73} 084009 (2006). arXiv:hep-th/0602269.

\bb{Wang2}S.~Wang,
\emph{Thermodynamics of Schwarzschild de Sitter spacetimes: Variable cosmological constant},
arXiv:gr-qc/0606109.

\bb{nuc}G.L.~Cardoso and V.~Grass,
\emph{On five-dimensional non-extremal charged black holes and FRW cosmology},
Nucl. Phys. B \textbf{803} 209 (2008). arXiv:0803.2819 [hep-th].

\bb{ex1}D.~Kastor, S.~Ray, and J.~Traschen,
\emph{Enthalpy and the Mechanics of AdS Black Holes},
Class. Quantum Grav. \textbf{26} 195011 (2009). arXiv:0904.2765 [hep-th].

\bb{Dolan2}B.P.~Dolan,
\emph{Pressure and volume in the first law of black hole thermodynamics},
Class. Quantum Grav. \textbf{28} 235017 (2011). arXiv:1106.6260 [gr-qc].

\bibitem{Dolan}B.P.~Dolan, ``Where is the PdV in the First Law of Black Hole Thermodynamics?" in \textit{Open Questions in Cosmology}, edited by G.J.~Olmo (\href{http://www.intechopen.com/books/open-questions-in-cosmology}{InTech}, 2012), \href{http://cdn.intechopen.com/pdfs/41690.pdf}{ch. 12}.

\bb{iso}M.~Cveti\v{c}, G.W.~Gibbons, D.~Kubiz\v{n}\'{a}k, and C.N.~Pope,
\emph{Black Hole Enthalpy and an Entropy Inequality for the Thermodynamic Volume},
Phys. Rev. D \textbf{84} 024037 (2011). arXiv:1012.2888 [hep-th].

\bibitem{Mann1}D.~Kubiz\v{n}\'{a}k and R.B.~Mann,
\emph{$P-V$ criticality of charged AdS black holes},
JHEP \textbf{07} 033 (2012). arXiv:1205.0559 [hep-th].

\bibitem{Mann2}S.~Gunasekaran, R.B.~Mann, and D.~Kubiz\v{n}\'{a}k,
\emph{Extended phase space thermodynamics for charged and rotating black holes and Born-Infeld vacuum polarization},
JHEP \textbf{11} 110 (2012). arXiv:1205.0559 [hep-th].

\bibitem{Wu}X.N.~Wu,
\emph{Multicritical phenomena of Reissner-Nordstr\"om anti–de Sitter black holes},
Phys. Rev. D {\bf 62} 124023 (2000).

\bb{br1}D.~Birmingham,
\emph{Topological black holes in anti-de Sitter space},
Class. Quantum Grav. \textbf{16} 1197 (1999). hep-th/9808032.

\bibitem{Emp}R.~Emparan,
\emph{AdS/CFT duals of topological black holes and the entropy of zero energy states},
JHEP \textbf{06} 036 (1999). hep-th/9906040.

\bibitem{Emp2}A.~Chamblin, R.~Emparan, C.V.~Johnson, and R.C.~Myers,
\emph{Holography, thermodynamics and fluctuations of charged AdS black holes},
Phys. Rev. D {\bf 60} 104026 (1999). hep-th/9904197.

\bibitem{Sahay}A.~Sahay, T.~Sarkar, and G.~Sengupta,
\emph{On the thermodynamic geometry and critical phenomena of AdS black holes},
JHEP \textbf{07} 082 (2010). arXiv:1004.1625 [hep-th].

\bibitem{Zhao}R.~Zhao, H.-H.~Zhao, M.-S.~Ma, and L.-C.~Zhang,
\emph{On the critical phenomena and thermodynamics of charged topological dilaton AdS black holes},
Eur. Phys. J. C  \textbf{73} 2645 (2014). arXiv:1305.3725 [gr-qc].

\bibitem{Wei}S.-W.~Wei snd Y.-X.~Liu,
\emph{Critical phenomena and thermodynamic geometry of charged Gauss-Bonnet AdS black holes},
Phys. Rev. D {\bf 87} 044014 (2013). arXiv:1209.1707 [gr-qc].

\bibitem{AdS}C.~Niu, Y.~Tian, and X.~Wu,
\emph{Critical Phenomena and Thermodynamic Geometry of RN-AdS Black Holes},
Phys. Rev. D {\bf 85} 024017 (2012). arXiv:1104.3066 [hep-th].

\bb{Maldacena2}E.~Papantonopoulos (Ed.), \emph{From Gravity to Thermal Gauge Theories: The AdS/CFT Correspondence}, Lecture Notes in Physics, Vol. 828, (Springer-Verlag, Berlin Heidelberg, 2011).

\bb{Maldacena3}S.S.~Gubser,
\emph{Breaking an Abelian gauge symmetry near a black hole horizon},
Phys. Rev. D \textbf{78} 065034 (2008). arXiv:0801.2977 [hep-th].

\bb{Maldacena4}S.A.~Hartnoll, C.P.~Herzog, and G.T.~Horowitz,
\emph{Building an AdS/CFT superconductor},
Phys. Rev. Lett. \textbf{101} 031601 (2008). arXiv:0803.3295 [hep-th].

\bb{Maldacena5}T.~Morita,
\emph{What is the gravity dual of the confinement/deconfinement transition in holographic QCD?},
Journal of Physics: Conference Series \textbf{343} 012079 (2012). arXiv:1111.5190 [hep-th].

\bb{Maldacena6}R.G.~Cai and N.~Ohta,
\emph{Deconfinement transition of AdS/QCD at $\mathcal{O}(\al '^3)$},
Phys. Rev. D {\bf 76} 106001 (2007). arXiv:0707.2013 [hep-th].

\bb{Maldacena7}R.G.~Cai and R.Q.~Yang,
\emph{Paramagnetism-Ferromagnetism Phase Transition in a Dyonic Black Hole},
To appear in Phys. Rev. D as a rapid communication. arXiv:1404.2856 [hep-th].

\bibitem{CFT1}A.~Strominger,
\emph{The dS/CFT correspondence},
JHEP \textbf{10} 034 (2001). arXiv:hep-th/0106113.

\bibitem{CFT2}A.~Strominger,
\emph{Inflation and the dS/CFT Correspondence},
JHEP \textbf{11} 049 (2001). arXiv:hep-th/0110087.

\bb{CFT3}C.M.~Hull and R.R.~Khuri,
\emph{Worldvolume theories, holography, duality and time},
Nucl. Phys. B \textbf{575} 231 (2000). arXiv:hep-th/9911082.

\bb{CFT4}P.O.~Mazur and E.~Mottola,
\emph{Weyl cohomology and the effective action for conformal anomalies},
Phys. Rev. D \textbf{64} 104022 (2001). arXiv:hep-th/0106151.

\bibitem{AAR}M.~Azreg-A\"{\i}nou and M.E.~Rodrigues,
\emph{Thermodynamical, geometrical and Poincar\'e methods for charged black holes in presence of quintessence},
JHEP \textbf{09} 146 (2013). arXiv:1211.5909 [gr-qc].

\bibitem{Kis}V.V.~Kiselev,
\emph{Quintessence and black holes},
Class. Quantum Grav. \textbf{20} 1187 (2003). arXiv:0210040 [gr-qc].

\bibitem{liv}G.~Comp\`{e}re,
\emph{The Kerr/CFT Correspondence and its Extensions},
\href{http://www.livingreviews.org/lrr-2012-11}{Living Rev. Relativity} \textbf{15} 11 (2012).
arXiv:1203.3561 [hep-th].

\bibitem{Nariai1}H.~Nariai,
\emph{On a new cosmological solution to Einsteins field equations of gravity},
The Science Reports of the Tohoku University, Series I, No.1, (1951).
Gen. Relativ. Grav. \textbf{31} 963 (1999).

\bibitem{Nariai2}J.~Podolsky,
\emph{The structure of the extreme Schwarzschild-de Sitter space-time},
Gen. Relativ. Grav. \textbf{31} 1703 (1999). arXiv:9910029 [gr-qc].

\bibitem{Nariai3}F.~Beyer,
\emph{Non-genericity of the Nariai solutions: I. Asymptotics and spatially homogeneous perturbations},
Class. Quantum Grav. \textbf{26} 235015 (2009). arXiv:0902.2531 [gr-qc].

\bibitem{Nariai3b}G.J.~Galloway,
\emph{Cosmological Spacetimes with $\La >0$},
arXiv:gr-qc/0407100.

\bibitem{Nariai4}C.G.~B\"{o}hmer and K.~Vandersloot,
\emph{Loop quantum dynamics of the Schwarzschild interior},
Phys. Rev. D \textbf{76} 104030 (2007). arXiv:0709.2129 [gr-qc].

\bb{Nariai5}N.V.~Mitskievich, M.G.~Medina Guevara, and H.V.~Rodr\'{\i}guez, ``Nariai-Bertotti-Robinson spacetimes as a building material for one-way wormholes with horizons, but without singularity", in \textit{Proceedings of the 11th Marcel Grossmann Meeting on General Relativity}, edited by H.~Kleinert, R.T.~Jantzen, and R.~Ruffini (\href{http://www.worldscientific.com/worldscibooks/10.1142/6997#t=toc}{World Scientific}, Singapore, 2008), p. 2181. arXiv:0707.3193 [gr-qc].

\bibitem{Nariai6}M.~Fennen and D.~Giulini,
\emph{An exact static two-mass solution using Nariai spacetime},
arXiv:1408.2713 [gr-qc].

\bb{AD}L.F.~Abbott and S.~Deser,
\emph{Stability of gravity with a cosmological constant},
Nucl. Phys. B \textbf{195} 76 (1982).

\bb{BBM}V.~Balasubramanian, J.~de Boer, and D.~Minic
\emph{Mass, Entropy and Holography in Asymptotically de Sitter Spaces},
Phys. Rev. D \textbf{65} 123508 (2002). arXiv:hep-th/0110108.

\bb{Tt1}C.~Teitelboim, ``Gravitational Thermodynamics of Schwarzschild-de Sitter Space", in \textit{Proceedings of the 5th Francqui Colloquium ``Strings and Gravity: Tying the Forces Together"}, edited by M.~Henneaux and A.~Sevrin (De Boeck \& Larcier, Bruxelles, 2003), p. 291. arXiv:hep-th/0203258.

\bb{Tt2}A.~Gomberoff and C.~Teitelboim,
\emph{de Sitter Black Holes with Either of the Two Horizons as a Boundary},
Phys. Rev. D \textbf{67} 104024 (2003). arXiv:hep-th/0302204.

\bb{Pad1}T.~Padmanabhan,
\emph{Classical and quantum thermodynamics of horizons in spherically symmetric
spacetimes},
Class. Quant. Grav. \textbf{19} 5387 (2002). arXiv:gr-qc/0204019.

\bb{Pad2}T.~Padmanabhan,
\emph{The holography of gravity encoded in a relation between entropy, horizon area
and the action for gravity},
Gen. Relativ. Grav. \textbf{34} 2029 (2002). arXiv:gr-qc/0205090.

\bb{Pad3}T.R.~Choudhury and T.~Padmanabhan,
\emph{Concept of temperature in multi-horizon spacetimes: analysis of Schwarzschild–De Sitter metric},
Gen. Relativ. Grav. \textbf{39} 1789 (2007). arXiv:gr-qc/0404091.

\bibitem{cold}S.~Fernando,
\emph{Cold, ultracold and Nariai black holes with quintessence},
Gen. Relativ. Gravit. \textbf{45} 2053 (2013). arXiv:1401.0714 [gr-qc].

\bibitem{Ads}G-Q.~Li,
\emph{Effects of dark energy on \textit{P-V} criticality of charged Ads black holes},
Phys. Letts. B \textbf{735} 256 (2014). arXiv:1407.0011 [gr-qc].

\bibitem{SF}S.~Fernando,
\emph{Nariai black holes with quintessence},
arXiv:1408.5064v1 [gr-qc].

\bibitem{V1}A.~Beesham, S.G.~Ghosh,
\emph{Naked Singularities in the Charged Vaidya-de Sitter Spacetime},
Int. J. Mod. Phys. D \textbf{12} 801 (2003). arXiv:gr-qc/0003044.

\bibitem{V2}H-Bin.~Sun, F.~He, and H.~Huang,
\emph{Statistical Entropy of Vaidya-de Sitter Black Hole to All Orders in Planck Length},
Int. J. Theor. Phys. \textbf{51} 1762 (2012).

\bibitem{GH} G.W.~Gibbons and S.W.~Hawking,
\emph{Cosmological event horizons, thermodynamics, and particle creation},
Phys. Rev. D \textbf{15} 2738 (1977).

\bibitem{irina} I.~Dymnikova and M.~Korpusik,
\emph{Thermodynamics of Regular Cosmological Black Holes with the de Sitter Interior},
Entropy \textbf{13} 1967  (2011).

\bibitem{MTZ} C.~Mart\'{\i}nez, R.~Troncoso, and J.~Zanelli,
\emph{de Sitter black hole with a conformally coupled scalar field in four dimensions},
Phys. Rev. D \textbf{67} 024008 (2003). arXiv:hep-th/0205319.

\bb{C5}E.~Winstanley,
\emph{Classical and thermodynamical aspects of black holes with conformally coupled scalar field hair},
Conf. Proc. C \href{http://inspirehep.net/search?p=find+EPRINT+gr-qc/0408046}{\textbf{0405132}} 305 (2004) (\href{http://www2.mate.polimi.it/ocs/viewpaper.php?id=36&cf=6}{Eds.} L.~Fatibene, M.~Francaviglia, R.~Giamb\`{o}, and G.~Magli).
arXiv:0408046 [gr-qc].

\bb{on-gtd}M.~Azreg-A\"{\i}nou,
\emph{Geometrothermodynamics: comments, criticisms, and support},
Eur. Phys. J. C \textbf{74} 2930 (2014). arXiv:1311.6963.

\bb{new}M.~Azreg-A\"{\i}nou,
\emph{Enthalpy, internal energy and thermodynamic volume of static possibly-variable-mass charged black holes},
in preparation.


\end{thebibliography}
\end{document}